\documentclass[letters, fleqn, usenatbib]{mnras}
\usepackage{amssymb,amsmath}
\usepackage{graphicx}
\usepackage{xcolor}
\usepackage{multirow}
\usepackage[T1]{fontenc}
\usepackage{ae,aecompl}
\usepackage{newtxtext,newtxmath}
\usepackage{longtable,listings}

\def\oc3{[O~{\sc iii}]$_c$}
\def\ob3{[O~{\sc iii}]$_b$}
\def\obj{SDSS J0141}
\def\dif{\mathop{}\!\mathrm{d}}

\title[A new TDE candidate at z=1.06]
{A new candidate for central tidal disruption event in SDSS J014124+010306 with broad Mg~{\sc ii} line at $z=1.06$}
\author[Zhang]
{Xue-Guang Zhang$^*$ \\
    School of physical science and technology, GuangXi University, No. 100, Daxue Road, 530004, Nanning, P. R. China}

\pubyear{2022}

\begin{document}

\label{firstpage}

\pagerange{\pageref{firstpage}--\pageref{lastpage}}

\maketitle

\begin{abstract} 
      In the Letter, a new candidate for central tidal disruption event (TDE) is reported in SDSS J014124+010306 
(=SDSS J0141) with broad Mg~{\sc ii} line at redshift $z=1.06$. Based on long-term photometric $ugriz$-band 
variabilities from SDSS Stripe82 Database and PHOTOOBJALL database, a central TDE is preferred with a 
1.3${\rm M_\odot}$ main-sequence star tidally disrupted by central black hole (BH) of $(14\pm2)\times10^6{\rm M_\odot}$ 
in SDSS J0141. Moreover, CAR process has been applied to confirm that the probability is only about 0.4\% that 
the long-term variabilities in SDSS J0141 are not related to TDE but from intrinsic AGN activities. Meanwhile, 
based on the apparent broad Mg~{\sc ii} emission lines, virial BH mass can be estimated as 
$245\times10^6{\rm M_\odot}$, 18 times larger than the TDE-model determined BH mass, providing further clues 
to support a central TDE in SDSS J0141, similar as the case in the TDE candidate SDSS J0159 with virial BH mass 
two magnitudes larger than M-sigma relation expected BH mass. Among the reported optical TDE candidates, SDSS 
J0141 is the candidate at the highest redshift. The results in the Letter indicate it should be common to detect 
TDE candidates in high redshift galaxies with broad Mg~{\sc ii} lines. 
\end{abstract}


\section{Introduction}

	TDEs (Tidal Disruption Events), indicators to massive black holes (BHs) and BH accreting systems, 
have been well studied in detail for more than four decades \citep{re88, lu97, gm06, gr13, gm14, mg19, st19, 
zl21} with accreting fallback debris from stars tidally disrupted by central black holes (BHs) leading to 
apparent time-dependent variabilities. More recent reviews on TDE can be found in \citet{gs21}. More recent 
two large samples of dozens of new TDE candidates can be found in \citet{vg21} and in \citet{sg21}.

	Among the reported TDE candidates, almost all the candidates are reported in inactive galaxies, 
such as the two candidates in non-active galaxies through the Stripe82 database by \citet{ve11}, the PS1-10jh 
and PS1-11af in inactive galaxies through the PanSTARRS (panoramic survey telescope and rapid response 
system) by \citet{gs12, ch14}, the iPTF16fnl in E+A galaxy through the PTF (palomar transient factory) by 
\citet{bl17}, the OGLE17aaj through the Optical Gravitational Lensing Experiment (OGLE) in quite weakly 
active galaxy by \citet{gr19}, and the well-known ASASSN-14ae, ASASSN-14li and ASASSN-19dj in nearby 
quiescent galaxies through the ASAS-SN (all-sky automated survey for supernovae) by \citet{ht14, ht16, 
hi21}, etc.. Besides the TDE candidates in quiescent galaxies, there are only a few TDE candidates reported 
in active galaxies, such as the SDSS J0159 \citep{md15}, CNSS J0019+00 \citep{an20}. However, in SDSS J0159, 
due to quite different virial BH mass from the M-sigma relation expected BH mass as discussed in \citet{zh19}, 
the broad Balmer emission lines are expected to be totally related to central TDE debris, indicating SDSS 
J0159 is not a normal broad line AGN (BLAGN). And in CNSS J0019+00, there are no broad emission lines, 
indicating CNSS J0019+00 is not a normal broad line AGN. It is an interesting objective to detect TDE 
candidates in normal broad line galaxies.

\begin{figure*}
\centering\includegraphics[width = 18cm,height=7cm]{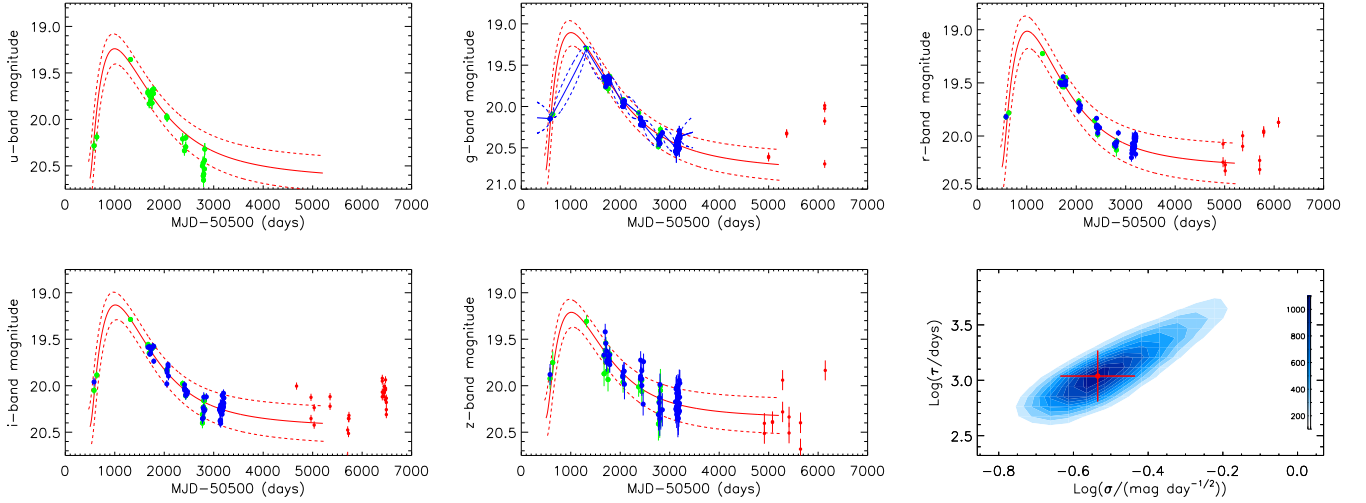}
\caption{Top panels and bottom left two panels show the SDSS $ugriz$-band light curves and the TDE 
model determined best descriptions. In each panel, solid blue and green circles represent the data points 
from the Stripe82 database and from the PHOTOOBJALL database, respectively. In each panel, solid and dashed 
red lines show the TDE model determined best descriptions and the corresponding confidence bands determined 
by uncertainties of the model parameters. In top middle panel, solid and dashed blue lines show the DRW 
process determined best descriptions to the $g$-band light curve and the corresponding 1sigma confidence 
bands. In each panel, sold circles plus error bars in red show the data points from the PanSTARRS, 
without considerations of magnitude difference between SDSS and PanSTARRS. Bottom right panel shows the 
two dimensional posterior distributions of the DRW process parameters of $\sigma$ and $\tau$ with solid 
circle plus error bars in red marking the accepted values.}
\label{lmc}
\end{figure*}

	Among the reported optical TDE candidates, strong broad Balmer and Helium emission lines are fundamental 
spectroscopic characteristics. And the reported broad emission lines can be related to disk-like structures from 
TDE debris, such as SDSS J0159 in \citet{md15, zh21}, ASASSN-14li in \citet{ht16}, PTF09djl in \citet{lz17}, 
PS18kh in \citet{ht19}, AT2018hyz in \citet{sn20, hf20}, etc., indicating the reported broad emission lines in 
the TDE candidates are not related to normal BLRs in normal BLAGN. Moreover, there are several TDE candidates, 
their UV-band spectra have been well checked, such as the PS18kh, ASASSN-15lh, ASASSN-14li, etc., there are no 
broad Mg~{\sc ii}$\lambda2800$\AA~ emission lines. In other words, if there was broad Mg~{\sc ii} emission line 
in UV spectra of a TDE candidate, the host galaxy of the TDE candidate could be probably a normal BLAGN. 
Therefore, to detect and report a TDE candidate with apparent broad Mg~{\sc ii} line is the main objective of 
the Letter.

	Among the high redshift objects in SDSS covering broad Mg~{\sc ii} emissions, a new TDE candidate in 
J014124+010306 (=SDSS J0141) at a redshift of 1.06 is reported in the Letter. Section 2 presents the long-term 
photometric SDSS $ugriz$-band variabilities of \obj. Section 3 shows the theoretical TDE model and fitting 
procedure applied. Section 4 shows spectroscopic properties and necessary discussions. Section 5 gives our 
final conclusions. We have adopted the cosmological parameters of $H_{0}=70{\rm km\cdot s}^{-1}{\rm Mpc}^{-1}$, 
$\Omega_{\Lambda}=0.7$ and $\Omega_{\rm m}=0.3$.

\section{Long-term Light curves in \obj}

	As a follow-up to our previous work on changing-look AGN in \citet{zh21b}, plans are underway for 
systematic searching changing-look AGN through multi-epoch SDSS spectra. When checking long-term variabilities 
of the candidates, the \obj~ with five repeated spectra is selected as the target of the Letter, due to its 
unique photometric variabilities.

	SDSS $ugriz$-band light curves of \obj~ are collected from the following two databases. First, the 
$griz$-band light curves are collected from the Stripe82 database \citep{bv08} with MJD from 51081 (September 
1998) to 53705 (December 2005). There are no $u$-band data points provided by the Stripe82 database. Second, 
the $ugriz$-band variabilities are collected from the SDSS PHOTOOBJALL database according to THINGID=120118318 
and the corresponding 29 photometric objids of \obj, by the following query applied in the SQL search tool in 
SDSS DR16,
\begin{lstlisting}
SELECT mjd, psfmag_g, psfmag_u, psfmag_r, psfmag_i, psfmag_z, 
  psfmagerr_g, psfmagerr_u, psfmagerr_r, psfmagerr_i, psfmagerr_z
FROM PHOTOOBJALL
WHERE objid = 1237657072231972919 or
  objid = 1237657192526905410 or objid = 1237657235444138067 or
  objid = 1237657364307247212 or objid = 1237657587628048446 or
  objid = 1237657737952297019 or objid = 1237657815264133192 or
  objid = 1237659915500454002 or objid = 1237660010022568005 or
  objid = 1237662969222070482 or objid = 1237663205462114377 or
  objid = 1237663506125488270 or objid = 1237663544780849275 or
  objid = 1237666302155292964 or objid = 1237666340803444954 or
  objid = 1237666383743484048 or objid = 1237666409527247014 or
  objid = 1237666499696328851 or objid = 1237666542643773631 or
  objid = 1237666637155991620 or objid = 1237666662926516356 or
  objid = 1237666727323500667 or objid = 1237646012704882989 or
  objid = 1237646648350802019 or objid = 1237653000602648648 or
  objid = 1237656513891467420 or objid = 1237656595492765803 or
  objid = 1237656909050544257 or objid = 1237656973454934076
\end{lstlisting}
Here, due to the point-like photometric images of \obj~ at redshift 1.06, PSF magnitudes are collected from the 
PHOTOOBJALL, rather than the Petrosian magnitudes commonly accepted for extended photometric images.

%

	The light curves are shown in Fig.~\ref{lmc}. The rise-to-peak and followed by smooth declining 
trend in each band light curve is apparent and can be well expected by a central TDE. Then, the theoretical 
TDE model can be considered to describe the variabilities of \obj.

\section{Theoretical TDE model applied to describe the Light Curves}

	More recent detailed descriptions on theoretical TDE model can be found in \citet{gr13, gm14, mg19} 
and in the corresponding public codes of TDEFIT and the MOSFIT. Here, based on the more recent discussions 
in \citet{mg19}, the theoretical TDE model can be applied by the following four steps, similar what we have 
done in \citet{zh22} to describe the X-ray variabilities in the TDE candidate {\it Swift} J2058.4+0516 with 
relativistic jet.

	First, standard templates of viscous-delayed accretion rates $\dot{M}_{at}$ are created, based on the 
TDEFIT/MOSFIT provided the $dm/de$ as distributions of debris mass $dm$ as a function of the specific binding 
energy $e$ after a star is disrupted, by the equations
\begin{equation}
\begin{split}
\dot{M}_{at}~&=~\frac{exp(-t/T_{v})}{T_{v}}\int_{0}^{t}exp(t'/T_{v})\dot{M}_{fbt}dt' \\
\dot{M}_{fbt}~&=~dm/de~\times~de/dt \ \ \ \ \ \ de/dt~=~\frac{(2~\pi~G~M_{\rm BH})^{2/3}}{3~t^{5/3}}
\end{split}
\end{equation}
where $M_{\rm BH}$ as the central BH mass, and $\dot{M}_{fbt}$ as the templates of fallback material rates 
created for standard case with central BH of $M_{\rm BH}=10^6{\rm M_\odot}$ and disrupted main-sequence star 
of $M_{*}=1{\rm M_\odot}$ and with a grid of the listed impact parameters $\beta_{temp}$ in \citet{gr13}, and 
$T_{v}$ as the viscous time after considering the viscous delay effects as discussed in \citet{gr13, mg19}. 
Here, a grid of 31 $\log(T_{v, temp}/{\rm years})$ range from -3 to 0 are applied to create templates of 
$\dot{M}_{at}$ for each impact parameter. Finally, templates of $\dot{M}_{at}$ include 736 (640) time-dependent 
viscous-delayed accretion rates for 31 different $T_{v}$ of each 23 (20) impact parameters for the main-sequence 
star with polytropic index $\gamma$ of 4/3 (5/3).

	Second, simple linear interpolations are applied to determine accretion rates $\dot{M}_{a}(T_{v},~\beta)$ 
for TDEs with input $\beta$ and $T_{v}$ different from the list values in $\beta_{temp}$ and in $T_{v, temp}$. 
Assuming $\beta_1$, $\beta_2$ in the $\beta_{temp}$ as the two values nearer to the input $\beta$ and $T_{v1}$, 
$T_{v2}$ in the $T_{v,temp}$ as the two values nearer to the input $T_{v}$, the first linear interpolation is 
applied to find the viscous-delayed accretion rates with input $T_{v}$ but with $\beta=\beta_1$ and 
$\beta=\beta_2$ by 
\begin{equation}
\begin{split}
\dot{M}_{a}(T_{v}, \beta_{1}) &= \dot{M}_{at}(T_{v1}, \beta_1) + \\
&\frac{T_{v}-T_{v1}}{T_{v2}-T_{v1}}(\dot{M}_{at}(T_{v2}, \beta_1) 
	- \dot{M}_{at}(T_{v1}, \beta_1))\\
\dot{M}_{a}(T_{v}, \beta_2) &= \dot{M}_{at}(T_{v1}, \beta_2) + \\
&\frac{T_{v}-T_{v1}}{T_{v2}-T_{v1}}(\dot{M}_{at}(T_{v2}, \beta_2)
	- \dot{M}_{at}(T_{v1}, \beta_2))
\end{split}
\end{equation}
Then, the second linear interpolation is applied to find the viscous-delayed accretion rates with input $T_{v}$ 
and with input $\beta$ by
\begin{equation}
\dot{M}_{a}(T_{v}, \beta) = \dot{M}_{a}(T_{v}, \beta_1) + \\
 	\frac{\beta-\beta_1}{\beta_2-\beta_1}(\dot{M}_{a}(T_{v}, \beta_2)
	- \dot{M}_{a}(T_{v}, \beta_1))
\end{equation}

	Third, for TDEs with input $M_{\rm BH}$ and $M_{*}$ different from $10^6{\rm M_\odot}$ 
and $1{\rm M_\odot}$, as discussed in \citet{gr13, mg19}, the actual viscous-delayed accretion rates 
$\dot{M}$ and the corresponding time information are created from the viscous-delayed accretion rates 
$\dot{M}_{a}(T_{v},~\beta)$ by the following scaling rations applied with input BH mass, mass and 
radius of the disrupted main-sequence star,
\begin{equation}
\begin{split}
&\dot{M} = M_{\rm BH,6}^{-0.5}\times M_{\star}^2\times
	R_{\star}^{-1.5}\times\dot{M}_{a}(T_{v}, \beta) \\
&t = (1+z)\times M_{\rm BH}^{0.5}\times M_{\star}^{-1}\times
	R_{\star}^{1.5} \times t_{a}(T_{v}, \beta)
\end{split}
\end{equation},
where $M_{\rm BH,6}$, $M_{\star}$, $R_{\star}$ and $z$ represent central BH mass in unit of ${\rm 10^6M_\odot}$, 
stellar mass in unit of ${\rm M_\odot}$, stellar radius in unit of ${\rm R_{\odot}}$ and redshift of host galaxy 
of a TDE, respectively. And the mass-radius relation discussed in \citet{tp96} has been accepted in the manuscript 
for main-sequence stars.

	Fourth, the time-dependent output emission spectrum in rest frame based on the TDE model expected accretion 
rate $\dot{M(t)}$ can be determined by the simple black-body photosphere model as discussed in \citet{gm14, mg19}, 
\begin{equation}
\begin{split}
&F_\lambda(t)=\frac{2\pi Gc^2}{\lambda^5}\frac{1}{exp(hc/(k\lambda T_p(t)))-1}(\frac{R_p(t)}{D})^2 \\
&R_p(t) = R_0\times a_p(\frac{\epsilon\dot{M(t)}c^2}{1.3\times10^{38}M_{\rm BH}/{\rm M_\odot}})^{l_p} \\
&T_p(t)=(\frac{\epsilon\dot{M(t)}c^2}{4\pi\sigma_{SB}R_p^2})^{1/4} \ \ \ \ \ a_p = (G M_{\rm BH}\times (\frac{t_p}{\pi})^2)^{1/3}	
\end{split}
\end{equation}
where $D$ means the distance to the earth calculated by redshift $z$, $k$ is the Boltzmann constant, $T_p(t)$ 
and $R_p(t)$ represent the time-dependent effective temperature and radius of the photosphere, respectively, 
and $\epsilon$ is the energy transfer efficiency smaller than 0.4, $\sigma_{SB}$ is the Stefan-Boltzmann constant, 
and $t_p$ is the time information of the peak accretion. Then, time-dependent apparent SDSS $ugriz$-band magnitudes 
$mag_{u,~g,~r,~i,~z}(t)$ can be well determined through the $F_\lambda(t)$ in observer frame convoluted with the 
accepted transmission curves of the SDSS $ugriz$ filters.

	Based on the four steps above, TDE model expected time dependent apparent $ugriz$-band magnitudes 
$mag_{u,~g,~r,~i,~z}(t)$ can be well created with seven parameters (redshift $z=1.0624$ accepted to \obj) of 
central BH mass $M_{\rm BH}$, mass $M_{\star}$ and polytropic index $\gamma$ (4/3 or 3/5) of the disrupted 
main-sequence star, the impact parameter $\beta$, the viscous-delay time $T_{v}$, the energy transfer efficiency 
$\epsilon$, and the two parameters of $R_0$ and $l_p$ related to the black-body photosphere. Moreover, there 
are five additional parameters $mag_{0}(u,~g,~r,~i,~z)$, applied to determine contributions of host galaxies 
(the none-variability components included in the light curves) to observed variabilities.

	Finally, through the Levenberg-Marquardt least-squares minimization technique (the MPFIT package) 
\citep{mc09}, the theoretical TDE model can be well applied to describe the SDSS $ugriz$-band light curves. 
Meanwhile, when the fitting procedure above is applied, there is one limitation to the model parameters. 
For an available TDE, the determined tidal disruption radius is limited to be larger than event 
horizon of central BH. Then, the determined TDE model parameters (with $\gamma=5/3$) and the corresponding 
uncertainties (1sigma errors computed from the covariance matrix) are: $\log(M_{\rm BH,6})\sim1.16\pm0.05$, 
$\log(M_\star/{\rm M_\odot})\sim-0.31\pm0.04$, $\log(\beta)\sim0.39\pm0.01$, $\log(T_{v})\sim-0.81\pm0.05$, 
$\log(\epsilon)\sim-0.55\pm0.05$, $\log(R_0)\sim-0.41\pm0.06$, $\log(l_p)\sim-0.14\pm0.06$, 
$\log(mag_0(u))\sim1.32\pm0.01$, $\log(mag_0(g))\sim1.32\pm0.01$, $\log(mag_0(r))\sim1.31\pm0.01$, 
$\log(mag_0(i))\sim1.31\pm0.01$, $\log(mag_0(z))\sim1.31\pm0.01$. Fig.~\ref{lmc} shows the TDE model determined 
best-fitting results and the corresponding confidence bands by uncertainties of the model parameters. 

	Before the end of the section, three points are noted. First, as discussed in \citet{rs18}, \obj~ 
has been classified as an extreme variability quasar (EVQ). Based on the improved DRW (Dampled Random Walk) 
process in \citet{koz10, zk13}, the public JAVELIN (Just Another Vehicle for Estimating Lags In Nuclei) code 
can lead to best descriptions to the light curves of \obj. Here, in top middle panel of Fig.~\ref{lmc}, the 
JAVELIN code determined best descriptions and corresponding confidence bands are shown as solid and dashed 
blue lines. Meanwhile, through the MCMC (Markov Chain Monte Carlo, \citet{fh13}) analysis with the uniform 
logarithmic priors of the DRW process parameters of $\tau$ and $\sigma$ ($SF_\infty\sim\sigma\sqrt{\tau}$ 
as the parameter used in \citet{rs18}), bottom right panel of Fig.~\ref{lmc} shows the posterior distributions 
of $\sigma$ and $\tau$, with $\log(\tau/days)\sim3.04$ and $\log(\sigma/(mag/days^{0.5}))\sim-0.535$, 
leading to the determined $\log(SF_{\infty}/mag)=\sim0.98$ which is about 1magnitude larger 
than the shown results in Fig.~9\ in \citet{rs18}, indicating \obj~ should have unique variabilities among 
the EVQs.

\begin{figure}
\centering\includegraphics[width = 8cm,height=5cm]{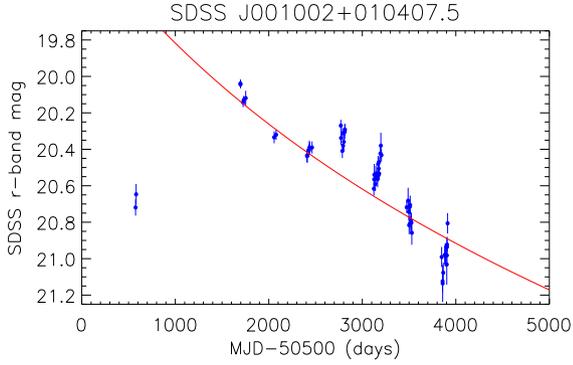}
\caption{The $r$-band light curve of SDSS J001002+010407.5 as a TDE candidate collected from S82Qs. 
Solid red line shows the trend described by $t^{-5/3}$ as TDE model expected.}
\label{fake}
\end{figure}

\begin{figure*}
\centering\includegraphics[width = 18cm,height=3.5cm]{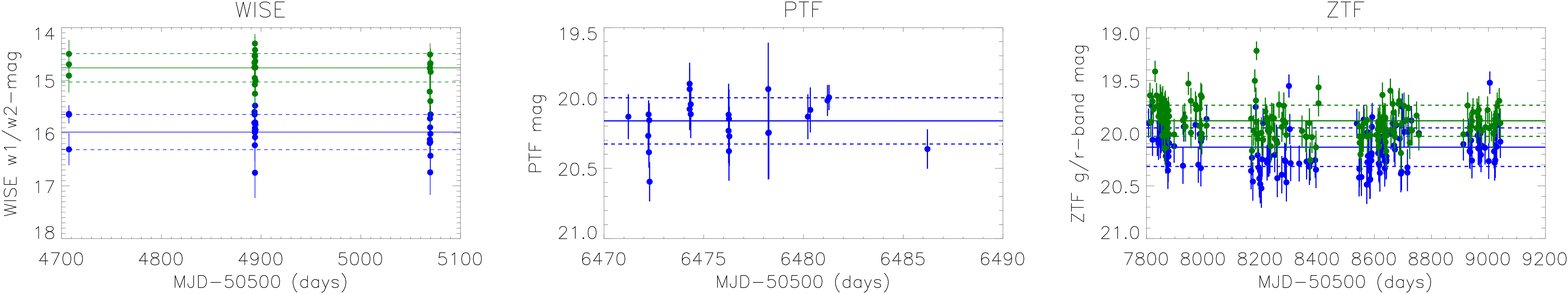}
\caption{The collected data points from WISE (w1 in blue, w2 in dark green), PTF and ZTF ($g$-band 
in blue, $r$-band in dark green). In each panel, horizontal solid line and dashed lines show the mean value and 
the range of standard deviation.
}
\label{zpw}
\end{figure*}

	Second, an interesting method is applied to check whether the shown variabilities  
in Fig.~\ref{lmc} is not from TDE but actually from intrinsic AGN activities. Based on the CAR (continuous 
Autoregressive) process preferred to describe AGN variabilities discussed in \citet{kbs09}:
\begin{equation}
\dif LMC_t=\frac{-1}{\tau}LMC_t\dif t+\sigma_*\sqrt{\dif t}\epsilon(t) + 19.77
\end{equation}
where $\epsilon(t)$ a white noise process with zero mean and variance equal to 1, and $19.77$ is the mean value 
of $LMC_t$ (the mean value of the SDSS $r$-band light curve of \obj)(different mean values in different 
bands have few effects on our following results). Then, a series of 1000 simulating light curves [$t_i$, $LMC_i$] 
can been created, with randomly selected values of $\tau$ from 50days to 5000days (similar as the reported values 
of extreme variability quasars in \citet{rs18}) and $\sigma_*$ leading the variance $\tau\sigma_*^2/2$ to be 0.095 
(the variance of SDSS $r$-band light curve of \obj), and $t_i$ are the same as the observational time information 
shown in Figure~\ref{lmc}. Among the 1000 simulating light curves, 4 light curves can be well described by 
theoretical TDE model, based on the criterion that the TDE-model determined best descriptions lead $\chi^2/Dof$ 
to be better than 3 (2.5 for the results shown in Fig.~\ref{lmc}). Therefore, the probability is only about 
0.4\% (4/1000) that the detected TDE-expected variabilities in \obj~ is mis-detected through the intrinsic 
AGN variabilities. Moreover, in order to confirm the probability of 0.4\%, all the 9258 quasars covered in the 
Stripe82 region (S82Qs) \citep{mi10} have their light curves been carefully checked. Among the 4763 S82Qs with 
light curves having number of available data points larger than 60, there are 11 quasars of which light curves 
have smooth declining trends $t^{\sim-5/3}$ as TDE model expected. The number ratio of TDE candidates among the 
S82Qs is about $11/4763\sim0.23\%$, to support the expected ratio of 0.4\% determined by the CAR simulating results. 
Fig.~\ref{fake} shows the light curve of one of the 11 TDE candidates among the S82Qs. Detailed discussions on 
the TDE candidates among the S82Qs are beyond scope of the Letter and will appear in our manuscript in preparation. 
Therefore, the TDE expected variabilities in \obj~ are confident enough.

	Third, besides the SDSS long-term variabilities, within searching radius smaller than 1\arcsec, 
long-term variabilities of SDSS J0141 have been collected from PanSTARRS with MJD from 55174 (Dec. 2009) 
to 56989 (Nov. 2014) shown as solid red circles in Fig.~\ref{lmc}. The PanSTARRS data points can be well 
followed by the TDE model, besides the PanSTARRS $g$-band data points, perhaps due to large magnitude 
difference relative to quite different $g$-filter transmission curves between SDSS and PanSTARRs. Moreover, 
within searching radius smaller than 5\arcsec, long-term variabilities can be collected from the PTF in 
Nov. 2014, from the WISE (Wide-field Infrared Survey Explorer) with MJD from 55207 (Jan. 2010) to 55570 (Jan. 
2011), and from the ZTF (Zwicky Transient Facility) with MJD from 58307 (Sep. 2018) to 59543 (Nov. 2021), 
shown in Fig.~\ref{zpw} with none apparent variabilities. The long-term none-variabilities not only can 
be well expected by TDE model at late times, but also can be accepted as indirect evidence to support 
that the variabilities shown in Fig.~\ref{lmc} are not from intrinsic AGN variabilities. Furthermore, 
based on the WISE data points shown in the left panel of Fig.~\ref{zpw}, the parameter $w1-w2$ is about 
1.22mag, a normal value applied to classify AGN by WISE colors \citep{as18}. Meanwhile, as the reported 
MIR flares related to TDEs in \citet{wy18}, WISE colors can be well around $w1-w2~\sim~1mag$ (see their 
table~1). Therefore, only based on the WISE colors, it is hard to provide further evidence to confirm 
that the variabilities in SDSS J0141 are not related to a central TDE but totally related to intrinsic 
AGN activities. However, based on the TDE model expected variabilities from SDSS and the more recent 
7.4years-long none-variabilities from PTF and ZTF, the central TDE is preferred in SDSS J0141.

\begin{figure}
\centering\includegraphics[width = 8cm,height=5cm]{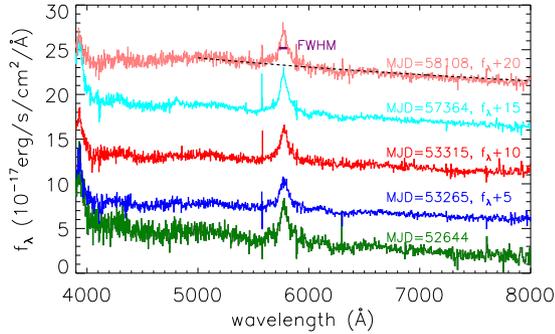}
\caption{The SDSS spectra in five different epochs. Dashed black line shows the power law described 
continuum emissions in the SDSS spectrum with MJD=58108. Horizontal purple line marks the width of FWHM 
of broad Mg~{\sc ii} in the SDSS spectrum with MJD=58108.}
\label{spec}
\end{figure}

\section{Virial BH mass from Spectroscopic Properties of \obj}

	SDSS spectra of \obj~ observed in MJD=52644, 53265, 53315, 57364, 58108 are shown in Fig.~\ref{spec} 
with broad Mg~{\sc ii} emission lines at $z=1.0624$. Moreover, for the spectrum with MJD=58108, continuum 
emissions can be described by $2.66\times\lambda^{-0.19}$, leading continuum luminosity at rest wavelength 
3100\AA~ to be about $\lambda L_{3100}\sim5.01\times10^{44}{\rm erg/s}$. 

	Through broad Mg~{\sc ii} emission lines, virial BH mass as discussed in 
\citet{pf04, sh11} can be estimated as 
\begin{equation}
\begin{split}
\log(\frac{M_{BH}}{{\rm M_\odot}})~&=~0.86~+~0.5\log(\frac{\lambda L_{3100}}{{\rm 10^{44}erg/s}})~+~
	2\log(\frac{FWHM}{{\rm km/s}})\\
&\ \ \ \ \ ~\sim~8.39
\end{split}
\end{equation}
with $FWHM\sim3900{\rm km/s}$ measured through the definition of Full Width at Half Maximum (FWHM) 
for the broad Mg~{\sc ii} from the spectrum observed in MJD=58108 after subtraction of the power law continuum 
emissions. Based on the results in Fig.~\ref{lmc}, the spectrum with MJD=58108 has few effects of the central 
TDE. The estimated virial BH mass is about 18 times larger than the TDE-model determined BH mass, indicating 
apparent contributions of TDE fallback accreting debris to broad Mg~{\sc ii} emission clouds. The distance of 
the broad Mg~{\sc ii} emission clouds to central BH could be quite smaller than the Virialization assumption 
expected BLRs sizes, leading to quite larger Virial BH mass in \obj, similar as what have been discussed in 
SDSS J0159 in \citet{zh19, zh21}.

	Actually, it is still an open question whether the TDE model determined BH masses are well consistent 
with intrinsic BH masses in TDE candidates. However, \citet{gm14, mg19, ry20, zl21} have shown that the 
long-term TDE model expected variabilities can be well applied to estimate central BH masses of TDE 
candidates. Meanwhile, for \obj~ at redshift 1.06, there is no way to measured stellar velocity dispersion in SDSS 
spectra, indicating that it is hard to measure more reliable BH mass through the M-sigma relation \citep{fm00, 
ge00, kh13} in \obj. Therefore, in the Letter, TDE model determined BH mass and Virial BH mass are compared in 
SDSS J0141. Certainly, further efforts are necessary to determine accurate central BH mass which will provide 
further clues to support or to against the central expected TDE in the \obj.

\section{Conclusions}

	Finally, we give our main conclusions as follows. A preferred TDE can be well applied to describe 
the long-term SDSS $ugriz$-band variabilities over 8 years in the \obj~ with apparent broad Mg~{\sc ii} 
emission lines at $z=1.06$, leading the TDE model determined BH mass to be about $(14\pm2)\times10^6{\rm M_\odot}$. 
Moreover, based on CAR process created artificial light curves, the probability is only about 0.4\% that the 
long-term variabilities in SDSS J0141 are from central AGN activities but not from a central TDE. Meanwhile, 
through the broad Mg~{\sc ii} emission lines, the virial BH mass can be estimated to be about 
$245\times10^6{\rm M_\odot}$, about 18times larger than the TDE-model determined BH mass, providing further 
clues to support the central TDE in \obj. Among the reported optical TDE candidates, SDSS J0141 is the 
candidate with the highest redshift. Moreover, it is feasible to detect more TDE candidates in galaxies 
with broad emission lines, not only broad Balmer lines and Helium lines but also broad Mg~{\sc ii} lines.

\section*{Acknowledgements}
Zhang gratefully acknowledges the anonymous referee for giving us constructive comments and 
suggestions to greatly improve our paper. Zhang gratefully acknowledges the grant support from NSFC-12173020. 
This Letter has made use of the data from the SDSS projects managed by the 
Astrophysical Research Consortium for the Participating Institutions of the SDSS-III Collaboration. The letter 
has made use of the public code of TDEFIT and MOSFIT, and MPFIT. This Letter has made use of the data 
from PanSRARRS, WISE, PTF and ZTF. 

\section*{Data Availability}
The data underlying this article will be shared on reasonable request to the corresponding author
(\href{mailto:aexueguang@qq.com}{aexueguang@qq.com}).

\label{lastpage}
\end{document}